\documentclass[twocolumn,showpacs,preprintnumbers,amsmath,amssymb]{revtex4}

\begin{document}
\draft

\title{FRACTIONAL PERIODICITY OF PERSISTENT CURRENTS:
A SIGNATURE OF BROKEN INTERNAL SYMMETRY}

\author{P. Singha Deo$^1$, P. Koskinen$^2$, M. Koskinen$^2$, 
and M. Manninen$^2$}
\address{$^1$ S. N. Bose National Centre for Basic Sciences, JD Block,
Sector III, Salt Lake City, Kolkata 98, India.}
\address{$^2$ Department of Physics, University of Jyv\"askyl\"a,
FIN-40351 Jyv\"askyl\"a, Finland}
\date{\today}

\begin{abstract}
We show from the symmetries of the many body Hamiltonian, cast
into the form of the Heisenberg (spin) Hamiltonian, that the fractional
periodicities of persistent currents are due to the breakdown
of internal symmetry and the spin Hamiltonian
holds the explanation to this transition. 
Numerical diagonalizations are performed
to show this explicitely.
Persistent currents therefore, provide an easy way to experimentally
verify broken internal symmetry in electronic systems.
\end{abstract}


\maketitle

Remarkable advances in fabrication techniques, now make it possible
to confine a few electrons in a conducting wire where electron motion
is governed by quantum mechanics, rather than classical mechanics.
The system becomes an electron waveguide within the confinement
potential. The phase coherence of the electrons is maintained over the
sample, making it possible to observe several intrinsic quantum
mechanical phenomena including
Aharonov-Bohm oscillations, universal conductance fluctuations,
quantized conductance in point contact, quenching of Hall resistance
in narrow cross \cite{boo}, 
current magnification effect \cite{deo1}, etc. A remarkable consequence
of such coherence is the existence of equilibrium persistent
currents \cite{hun}
in a ring threaded by an Aharonov-Bohm flux,
that was first predicted theoretically \cite{but}
and subsequently detected experimentally \cite{exp} in mesoscopic
systems. Aharonov-Bohm flux here refers to a situation
when the magnetic field is restricted to a small region in the
center of the ring, and the electrons in the ring do not feel
the magnetic field. Although these are
equilibrium currents, they can yield information about transport
\cite{deo2}
and may help us to understand the effects of electron-electron
interaction on transport, using Hamiltonian diagonalization techniques.

Interacting fermions exhibit very novel properties that fascinate
scientists for a long time. Interacting nucleons for example has
shown many novel features of interactions \cite{boh}. 
Of special relevance to this work, is the discovery that
certain heavy nuclei can exhibit rotational excitations,
that could not be explained by the shell structure of a spherical
neuclous. Initial understanding was provided by Bohr and Mottelson,
in terms of collective modes of a deformed nuclei, in the simplest
model as rigid rotation of the deformed nucleus.
Similar ideas of {\it internal} symmetry breaking,
explained the details of the mass spectra of alkali metal 
clusters\cite{clemenger85}, and suggest an existence of static
spin-density waves in quantum dots\cite{kos1}.
Indeed, mesoscopic systems give us an unique opportunity
to access regimes that do not occur naturally and study
a few electrons in man made quantum dots, both 
experimentally and
simultaneously with almost 
exact theoretical methods. Hence these systems can give us a rare
opportunity to study how few electron properties evolve into
macroscopic collective properties as we increase the number of electrons.

Wigner crystalization of electrons is one such bulk phenomenon
and is still a very debatable
issue, although it was first proposed a very long time ago. 
We hereby exclude the situation when the quantum mechanical
kinetic energy or uncertainty of an electron can be quenched
by a strong magnetic field.
Theoretically one can find signatures of a crystal structure when one looks
at the conditional probability (the probability of
finding other electrons when the coordinates of one electron
is fixed by hand) of the interacting electrons, while the
probability itself (density) does not show any sign of a crystal structure.
The conditional probability shows oscillations\cite{borrmann01}
which suggest broken internal symmetry. However, in finite systems
the situation is not so clear due to the fact that the 
correlation coming from the Pauli exclusion principle alone
will cause oscillating conditional probability at short distances,
even in noninteracting systems.
On the other hand a recent work of Koskinen et al \cite{kos1}
raises the
issue that a few electron system in a quantum dot
and quantum ring can exhibit broken internal
symmetry.  Their mean field studies
of the electron probability (density) showed a perfect crystal structure.
It was subsequently shown \cite{vei} that the quantum 
mechanical superposition of states
is destroyed by the non-linearity of the mean field and
as a result the internal symmetry
is mapped out (as a consequence the
angular momentum quantum number does not take
integral values anymore). 
Effects of non-linearity
cannot be ignored (the exact cause of non-linearity
not being important) and can well lead to
a quenching of the
quantum mechanical uncertainty and result in Wigner crystals. 
Hence at this state it would be useful
to find some experimental ways of determining if the internal symmetry
is broken or not and what are the signatures that one
should look for when treating larger systems using
approximate methods like mean field theories and 
effective Hamiltonians. For example, in nuclei, the experimental evidence
of such broken symmetry states can be obtained from the
rotational spectrum.  
The purpose of this work is to show that the measurement of the 
persistent current can give signature of broken internal symmetry
of electronic states in small quantum rings.

The many-body Hamiltonian for electrons in a quasi-1D-ring
can be written as 
\begin{equation}
H=\sum_{i}\left(-{{\hbar}^2\over{2m^*}}\nabla_i^2 + V(r_i)\right)
+\sum_{i<j}{{e^2}\over{4\pi\epsilon_0\epsilon\vert {\vec r}_i
-{\vec r}_j\vert}}
\label{hamilton}
\end{equation}
where $V$ is the potential confining the electron in the ring.
$V$ is assumed to have a circular symmetry.
Koskinen {\it et al} \cite{kos2,koskinen2002}
have performed exact numerical computations
for a few electrons confined in such a ring and shown that
for narrow rings the many-body spectrum can be described
essentially exactly with a simple model Hamiltonian
\begin{equation}
H_{eff}= J\sum_i {\vec S}_i\cdot {\vec S}_j +
{{1}\over{2I}}M^2 + \sum_\alpha \hbar\omega_\alpha n_\alpha
\label{modelh}
\end{equation}
where the first term is an antiferromagnetic Heisenberg 
Hamiltonian, the second term describes rigid rotations of 
the electron system ($M$ is the centre of mass angular momentum
and $I$ is the moment of inertia), and
the last term describes the vibrational states of localized
electrons. Koskinen {\it et al} \cite{kos2,koskinen2002}
compared the energy spectra
of exact diagonalization of Hamiltonian (\ref{hamilton}) to those
of the model Hamiltonian (\ref{modelh}) and found
an excellent agreement for hundreds of many-body states in rings with
from 2 to 7 electrons.

The model Hamiltonian can be understood as a result of localization
of electrons and forming a Wigner molecule which is freely
rotating in the external potential\cite{kos2}.
Related ideas of electron localization in noncircular dots
had been suggested earlier\cite{jef,cre}.
Assuming localization, the Coulomb energy of the 
exact Hamiltonian (\ref{hamilton}) can be expanded around the 
classical equilibrium positions of electrons.
This leads to potential wells at each classical site. 
The tight binding model of the system can be described by a 
half filled Hubbard model, which in the limit of the strong
Coulomb energy (Hubbard $U$) leads to the antiferromagnetic
Heisenberg model \cite{vol}. 
It is important to note, however, that for our continuous system 
where the localizing potential
is not an external potential, we do not get an insulating
phase for the half filled case \cite{kus}.
The rigid rotation and the the vibrations of the localized
electrons can be assumed to separate out from the spin Hamiltonian
leading to the simple effective Hamiltonian of Eq. (\ref{modelh}).
The antisymmetry of the total Hamiltonian have to be taken into 
account by matching the symmetries of the different parts of the 
wave function (spin, rotations, vibrations)\cite{koskinen2002}.
In the present case when we are studying only the ground state
properties (persistent current being a ground state
property) the vibrational states do not play any role.

An external magnetic field will bring two additional terms 
in the model Hamiltonian. The gauge field will change the 
angular momentum part and the direct interaction with the 
electron spins will add a Zeeman term. 
Since we are interested in the equilibrium persistent currents,
we will assume that the magnetic field is confined only
inside the ring so that the Zeeman energy is absent.
By ignoring the vibrations
(which have higher energy than rotations)
the model Hamiltonian in the presence of a magnetic field 
flux $\phi$ penetrating the ring is
\begin{equation}
H_B=J\sum_{i,j}{\vec S}_i\cdot {\vec S}_j 
+{{(M-N\phi)^2}\over{2I}}
\label{modelb}
\end{equation}
where $M$ is the angular momentum, $N$ the number of electrons in the ring and
$\phi$ is the flux through the ring in units of $\phi_0=hc/e$
i.e., $\phi$=${e \over hc} \int {\vec A}\cdot {\vec dr}$ 
$\vec A$ being the vector potential.
The strongly interacting case here correspond to $J \rightarrow 0$,
$J$=0 being the classical case when the electrons do not overlap. 
In this case, there is no
uncertainty in the internal frame of the system and it is a perfect
crystal in its internal frame.
It is important to notice that keeping $I$ fixed, the small $J$
limit correspond to a narrow ring with strong correlation and the
large $J$ limit approaches the non-interacting case.

The correspondence between (1) and (3) shown
in Ref. \cite{kos2} in the absence of flux,
can be easily extended to the case
when there is Aharonov-Bohm flux. First of all
the Hamiltonian in (3) cannot have any extra
contribution from the 2nd term in the Hamiltonian in (1).
This is because when we write the Hamiltonian in (1)
in terms of a centre of mass coordinate
and relative coordinates,
the second term contains only
relative coordinates,
and will only affect the vibrational states,
which in turn does not affect
the persistent currents. This means that $J$
is independent of flux because the second term in
(1) only depends on relative coordinates and is not
affected by the flux.
Another way to prove this is to show that the Coulomb
matrix elements are independent of Aharonov Bohm flux,
which can be shown analytically.
This is also evident in the numerical calculations \cite{pie,tap}.
Secondly, the flux also cannot affect the relative motions (kinetic
energy part) of the electrons and this was proved in Ref. \cite{wei}.
The proof essentially puts all the flux dependence on the wavefunction
by gauging away the flux dependence of the Hamiltonian. Writing the many
body wavefunction in presence of flux as a linear combination of Slater
determinants, constructed from the flux dependent, non-interacting,
single particle wave functions, it is easy to show that all the flux
dependence of the relative coordinates cancel each other. The second term
in (3) also directly follow from there, once we put the flux dependence
back into the Hamiltonian.

The persistent current can be determined as a derivative of the 
ground state energy with respect of the flux\cite{hun}.
Consequently, it is sufficient to study the periodicity of the 
ground state energy as a function of the flux.
First of all it can be shown that the Heisenberg Hamiltonian remains
unchanged under the transformation $M \rightarrow M+N$ and so we need to
consider only the first N eigen-energies of the system.
In Fig. 1 we show a contour plot of the eigen-energies for N=
4, 5, 6 and 8 in the $J-\phi$-plane.
When $J \rightarrow $0 
then the periodicity with flux is $\phi_0/N$ and correspond to
the case when the flux quantum is $\phi_0/N$ corresponding to the
rigid rotor of charge $Ne$. As $J$ is increased, signifying that
the electrons are getting delocalized and overlapping with each other,
the periodicity changes smoothly to $\phi_0$. For even number of
electrons, $\phi_0/N$
periodicity first changes over to $\phi_0/2$ periodicity before
the full $\phi_0$ periodicity is recovered. For odd $N$ also
$\phi_0$ periodicity changes to $\phi_0/2$
periodicity, which unlike the case of even $N$, 
remain all the way up to $J=\infty$. This can also be see in
Fig. 2 where we plot the $M$ values of the ground state for
different flux $\phi / \phi_0$
and $Jr_0^2$ ($r_0$ being the radius of the ring, $I=Nmr_0^2$). 
The ground state switches
its $M$ values as shown in the figure 2
in the different parameter regimes.
While for even $N$ one can see converging
phase regions that cannot be extended to infinity, 
for $N=5$ the line separating $M=1$ and $M=4$
is a vertical line that can be extended to infinity. This is also
consistent with the fact that when $J$ is large we should recover
the free electron results and odd number of spin-full free
electrons in a ring always give $\phi_0/2$ periodicity
\cite{los}. But,
for even $N$ the $\phi_0/N$ periodicity first changes to
$\phi_0/2$ periodicity as $J$ is increased, 
and then the free electron result
of $\phi_0$ periodicity is recovered for very large $J$. 

The value of flux which gives the minimum total energy
for large $J$ depends
on the number of electrons in the ring. For even particles with
$N=4$, 8, 12, etc. ($N/2$ is even) 
the minimum is at flux $\phi=\phi_0/2$ (see fig. 3) while
for $N=6$, 10 etc. ($N/2$ is odd) 
the minimum is at $\phi=0$. The reason is the
symmetry of the solution of the Heisenberg Hamiltonian
as can be proved for any $N$ from group theoretical analysis.
For odd $N/2$ the minimum energy corresponds 
to $M=0$ and the second lowest state
has $M=N/2$ while
for even $N/2$ the minimum energy has $M=N/2$ and the second lowest state
has $M=0$ (see fig. 3 as an example).
In both cases it happens that at a certain region of $J$, when the
angular momentum is increased, the ground state jumps between these
two lowest states leading to $\phi_0/2$ periodicity in the
total energy and persistent current. When $J$ becomes large enough
then the splitting between these two lowest states also become
very large and
only the ground state of the Heisenberg Hamiltonian matters
and $\phi_0$ periodicity is obtained.
For odd number of electrons (odd $N$)
the situation is different
since there are two angular momentum values corresponding
to the minimum energy of the Heisenberg Hamiltonian.
For noninteracting electrons these two values appear
at angular momenta (in the limit of $N\rightarrow \infty$)
$N/4$ and $3N/4$.
Consequently, there are two equal energy minima leading to
$\phi_0/2$ periodicity even at infinitely large values of $J$.
In the case of small number of electrons the energy minimum
can not occur exactly at $N/4$ and
the $\phi_0$ and $\phi_0/2$ periodities are
superimposed as in Fig. 1(b). This
signifies that for intermediate $J$ values, when there is
no longer any correlation between all the $N$ particles as
in a rigid rotor, 
there still seems to be correlation between electron pairs
leading to quasi-particles with charge $2e$.
For odd $N$ since every electron cannot find a pair (similar
to the parity effect in superconducting grains), such
correlation is not possible. Spin
values may or may not change as flux is changed from 0 to
$\phi_0$. The sequence of change is depicted in Fig. 2.
The $\phi_0/N$ periodicity occurs when $J$ is small and
the splitting between the states is so small that the flux
can create transition through all the states.

It should be noted that the flux dependence of the eigenenergies,
in a clean ring, should be the same in presence and absence of interactions
\cite{wei}. A many body eigenenergy, in absence of interactions, should
change parabolicaly with $N\phi$.
Our model Hamiltonian in (3) is consistent with this
and the flux dependence of a given state is always parabolic in the
calculated eigenenergies. However, as the periodicity is reduced by N,
the amplitude of the persistent currents is also reduced by N as
the Brillouin zone becomes 1/N of the non-interacting case. For large
$J$, when we recover the $\phi_0$ periodicity, once again the flux
dependence of a particular many body state changes parabolically with
$N\phi$, all the way up to the non-interacting zone boundary.

The relation of the parameters of the model Hamiltonian (2) to those of the 
original Hamiltonian (1) require exact diagonalization of the latter.
The results of Koskinen et al\cite{kos2,koskinen2002} indicate that the onset 
of the $\phi_0/N$ periodicity, which happens at $Jr_0^2\approx 0.1$,
can be obtained with a ring with $V(r)={1\over 2}m^*\omega_0^2(r-r_0)^2$,
by choosing (in effective atomic units) $\omega_0\approx 1/m^*r_0^2$ 
(for $N=6$).
For a material with $m^*=0.1$ and $\epsilon=10$ this condition could 
be achieved,
for example, with $r_0=80$ nm and $\omega_0=1.7$ meV.

Fractional periodicities of persistent currents in a 1D
Hubbard ring has been discussed before \cite {kus, wu, kot}.
These studies correspond to a situation when there is an
externally applied periodic potential or a lattice.
In presence of such an externally applied potential
the concept of Wigner crystal is not meaningful.
The observed fractional periodicities, Kotlyar et al\cite{kot} associate
with magnon excitations. 
How this interpretation relates to our findings 
is an interesting subject of future studies.

Breakdown of internal symmetry of a many body system \cite{kos1},
crystallization of electrons in the bulk (Wigner crystals),
fractional periodicities of persistent currents \cite{tap} are three
different intersting research topics that are brought together
in this work. It is shown that fractional periodicity of
persistent currents is due to the breakdown of internal
symmetry. 
The Hamiltonian,
diagonalized upto 8 electrons in a ring threaded by a flux
show this explicitely and symmetry considerations establish
this for any $N$. Broken internal symmetry in electronic
systems is of
special importance as the interaction between electrons
is well known as compared to that between neucleons.
It is difficult to observe such broken symmetry states
because most of the physical quantities that can be measured,
do not depend on whether the internal symmetry is
broken or not. In the nuclei, the only evidence of such
broken internal symmetry comes from the rotational and vibrational
spectrum of a nuclei. For electrons embedded in a solid
the equilibrium persistent currents provide a way
to find this evidence.
At present a few electron ring can be realized \cite{lor} and possibly
reveals the much sought experimental proof
of broken internal symmetry in an electronic system.

This work has been supported by the Academy of Finland under 
the Finnish Centre of Excellence Programme 2000-2005 (Project No.
44875, Nuclear and Condensed Matter Programme at JYFL).

\centerline{\bf Figure Captions}

\noindent Fig. 1. 2D plot of the energy of the Heisenberg Hamiltonian
for electrons in a 1D ring versus $Jr^2$ and $\phi/\phi_0$
for N=4 (Fig. a), 5 (Fig. b), 6  (Fig. c) and 8 (Fig. d). 
The dark areas are maxima and bright areas are minima. The persistent
is the derivative of the energy with respect to the flux and so
for the persistent currents dark areas are minima and bright areas
are maxima.

\noindent Fig. 2. Phase diagram of the ground state angular momentum
of the Heisenberg Hamiltonian
for electrons in a 1D ring versus $Jr^2$ and $\phi/\phi_0$
for N=4 (Fig. a), 5 (Fig. b), 6 (Fig. c) and 8 (Fig. d). 
The region with a particular shade denotes
the $Jr^2$ and $\phi/\phi_0$, corresponding to which the ground state
angular momentum is $M$, where $M$ is designated in the shaded regions
as M(S), where S is the total spin.

\noindent Fig. 3. States of an 8 electron system in a 1D ring as obtained
by diagonalizing the Heisenberg Hamiltonian. $\phi/\phi_0$=0,
$Jr^2=1.0$, the M values
are given at the base and the S values are labeled by the side of each state.
The difference between the lowest state and the highest state in the Fig.
is 5.651$Jr^2$.

\end{document}